# Network-Based Video Recommendation Using Viewing Patterns and Modularity Analysis: An Integrated Framework

**Mehrdad Maghsoudi[1], Mohammad Hossein Valikhani[2], and Mohammad Hossein Zohdi[2]**

[1] Department of Industrial and Information Management, Faculty of Management and Accounting, Shahid Beheshti University, Tehran, Iran

[2] Department of Management, Economics and Progress Engineering, Iran University of Science and Technology, Iran, Tehran, Iran

Corresponding author: Mehrdad Maghsoudi(e-mail: M_Maghsoudi@sbu.ac.ir )

**ABSTRACT** The proliferation of video-on-demand (VOD) services has led to a paradox of choice, overwhelming users with vast content libraries and revealing limitations in current recommender systems. This research introduces a novel approach by combining implicit user data, such as viewing percentages, with social network analysis to enhance personalization in VOD platforms. The methodology constructs user-item interaction graphs based on viewing patterns and applies centrality measures (degree, closeness, and betweenness) to identify important videos. Modularity-based clustering groups related content, enabling personalized recommendations. The system was evaluated on a documentary-focused VOD platform with 328 users over four months. Results showed significant improvements: a 63% increase in click-through rate (CTR), a 24% increase in view completion rate, and a 17% improvement in user satisfaction. The approach outperformed traditional methods like Naive Bayes and SVM. Future research should explore advanced techniques, such as matrix factorization models, graph neural networks, and hybrid approaches combining content-based and collaborative filtering. Additionally, incorporating temporal models and addressing scalability challenges for large-scale platforms are essential next steps. This study contributes to the state of the art by introducing modularity-based clustering and ego-centric ranking methods to enhance personalization in video recommendations. The findings suggest that integrating network-based features and implicit feedback can significantly improve user engagement, offering a cost-effective solution for VOD platforms to enhance recommendation quality.

## I. INTRODUCTION

The rapid proliferation of video-on-demand (VOD) services has revolutionized the entertainment landscape, providing viewers unparalleled access to vast content libraries.[1]. Major players like Netflix, Amazon Prime Video, and Hulu have transformed media consumption, catering to diverse preferences and interests [2, 3]. The expansion of subscription video-on-demand (SVOD) services has significantly impacted the entertainment technology industry since 2010. SVOD services require users to subscribe for access to premium video content. In Thailand, the SVOD market was valued at US$146 million in 2021, with an expected compound annual growth rate (CAGR) of 13.89% from 2021 to 2025, and an average revenue per user of US$30.78 [4]. This growth can be attributed to factors such as the ubiquity of high-speed internet, the convenience of on-demand viewing, and the popularity of binge-watching [5].

However, the abundance of available content on VOD platforms presents a paradox of choice, with many users feeling overwhelmed by the sheer volume of options [6]. This situation highlights a significant research gap: current recommender systems often fail to effectively manage the vastness of content, leading to user dissatisfaction and suboptimal engagement. Recommender systems have emerged as a critical tool for streaming services to alleviate this issue and assist users in discovering personally relevant content. These systems employ various algorithmic approaches, including content-based filtering [7, 8], collaborative filtering [8-11], and hybrid methods [12, 13], to generate personalized recommendations based on user preferences and behavior.

The primary objective of this research is to address the limitations of existing recommender systems by incorporating implicit user data and advanced social network analysis techniques. Existing recommender systems in the VOD domain have primarily relied on explicit user feedback, such as ratings and reviews [10, 13, 14], which may not fully capture the complexities of user engagement and satisfaction. Furthermore, evaluating these systems has often focused on metrics derived from user ratings on platforms like MovieLens [11, 15, 16] and IMDb [17, 18], neglecting the rich implicit data generated through user interactions with VOD services.

The main problem this research tackles is the insufficient utilization of implicit user data and the lack of integration between social network analysis and video similarity metrics in current recommender systems. While graph-based recommendation systems have been explored in previous studies [19], this research extends the application by combining advanced social network analysis techniques with video similarity graphs. The integration of centrality measures and modularity-based clustering for video recommendation offers a novel way to capture both user preferences and video network structure. [20], allowing for highly personalized and efficient content suggestions [21]. This integration proves particularly effective for video platforms where both online and offline social interactions play a role [22].

Centrality measures in network analysis provide distinct advantages for recommendation systems compared to deep learning approaches. While deep learning models can capture complex patterns, they often struggle with interpretability and cold-start problems [23, 24]. This research employs a network-based approach to address these limitations by providing transparent and interpretable recommendations derived from explicit network structures and user viewing patterns. The centrality metrics utilized in this study—degree, closeness, and betweenness—are grounded in robust theoretical foundations of social network analysis, ensuring that the recommendations are both explainable and computationally efficient.

Additionally, this research employs a hybrid approach combining social network analysis with implicit feedback, providing a robust framework that can handle both new users and new items through network structure, while still leveraging the rich patterns found in user viewing behavior [25]. This approach aligns with recent research suggesting that graph-based methods can provide comparable or superior performance to deep learning models in certain recommendation scenarios, particularly when interpretability and cold-start handling are prioritized [26, 27].

This research proposes a novel recommender system framework that leverages implicit user data and social network analysis techniques to enhance personalization in VOD platforms. By incorporating data on viewing histories, watch durations, social connections, and content engagement, this study aims to develop a more comprehensive understanding of user preferences and behavior. The proposed approach involves constructing user-item interaction graphs from VOD consumption patterns and applying graph-based algorithms, such as community detection and link prediction, to identify socially and behaviorally relevant content for each user.

The methodology employed in this research encompasses several key stages. First, extensive VOD consumption data is collected from a leading streaming platform, including user demographics, viewing histories, engagement metrics, and social connections. This data undergoes rigorous preprocessing and feature engineering to construct user-item interaction graphs that capture the intricate relationships between users and content. Graph-based algorithms, such as community detection and link prediction, are then applied to these graphs to identify content that aligns with users' social circles and behavioral patterns.

A combination of offline and online metrics is employed to evaluate the performance of the proposed recommender system. Offline evaluation involves traditional metrics such as precision [28-31], recall [29, 32], and normalized discounted cumulative gain (NDCG) [33-35], assessed through cross-validation of historical data. Online evaluation is conducted through a live user study, measuring metrics such as click-through rate (CTR) [36, 37], view completion rate [38], and user satisfaction ratings [39]. The effectiveness of the proposed approach is compared to existing rating-based methods and state-of-the-art recommender systems to demonstrate its superiority.

This research offers several significant contributions to the field of VOD recommender systems. By integrating implicit user data and social network analysis, the proposed framework captures nuanced user preferences and provides highly personalized recommendations. The findings of this study provide valuable insights for streaming platforms seeking to enhance user engagement, reduce churn, and improve overall user satisfaction. Moreover, the proposed approach has the potential to be generalizable to other domains involving user-item interactions, such as e-commerce and social media.

The remainder of this paper is organized as follows. Section 2 provides an overview of related work in the field of recommender systems, focusing on VOD platforms and the use of implicit user data and social network analysis. Section 3 describes the proposed methodology, including data collection, preprocessing, graph construction, and algorithmic approaches. Section 4 presents the experimental setup and evaluation metrics, while Section 5 discusses the results and comparative analysis. Finally, Section 6 concludes the paper and outlines future research directions.

## II. Literature Review

### A. Recommender systems

Recommender systems play a crucial role in providing personalized recommendations by analyzing user data and interactions with items [40]. These systems have become essential tools in various industries, aiding in decision-making processes and enhancing user experiences [41]. With the increasing volume of information available, recommender systems help alleviate the issue of information overload by offering tailored suggestions [42]. Among the various types of recommender systems, collaborative filtering stands out as a prominent algorithm that leverages data from multiple users to generate recommendations [43]. This method involves analyzing user behavior and preferences to make suggestions based on similarities with other users [40].

Collaborative filtering is widely recognized for its effectiveness in recommendation systems and is considered one of the most successful techniques in this field [44]. By accumulating user ratings and identifying commonalities among users, collaborative filtering can provide accurate and relevant recommendations [45]. This approach has been extensively studied since the 1990s and has significantly contributed to the advancement of recommendation system research [46]. Moreover, collaborative filtering algorithms are versatile and can be implemented in various domains, such as e-commerce, entertainment, and tourism [44].

In contrast to collaborative filtering, content-based filtering algorithms focus on the attributes of items and users' preferences to make recommendations [47]. These algorithms analyze the content of items to identify similarities and suggest relevant items to users based on their past interactions [48]. Content-based filtering is particularly useful in scenarios where user preferences are well-defined and explicit features of items are available for comparison [49]. By utilizing models that assess document similarities, content-based filtering can offer tailored recommendations that align with users' interests [50].

Another approach to recommendation systems is knowledge-based filtering, which relies on domain knowledge to make suggestions [51]. By understanding the characteristics of items and users, knowledge-based filtering algorithms can provide personalized recommendations that align with specific requirements or constraints [52]. This method is beneficial in situations where explicit knowledge about items is crucial for generating accurate recommendations [53]. Additionally, demographic-based filtering algorithms consider demographic information, such as age, gender, or location, to tailor recommendations to specific user segments [54]. By incorporating demographic data, these algorithms can enhance the relevance and effectiveness of recommendations for different user groups[55].

Combined filtering algorithms integrate multiple recommendation approaches, such as collaborative filtering, content-based filtering, and demographic filtering, to improve recommendation accuracy [56]. By leveraging the strengths of different algorithms, combined filtering approaches can offer more comprehensive and diverse recommendations to users [57]. These hybrid systems aim to overcome the limitations of individual algorithms and enhance the overall recommendation quality [58]. By combining various filtering techniques, hybrid systems can provide more accurate and diverse recommendations that cater to different user preferences and contexts [59].

### *B. Social Network Analysis*

Social network analysis (SNA) involves the examination of dynamic social structures composed of individuals or entities, known as "nodes," which are connected by various forms of interdependencies [60]. These interdependencies may include friendships, kinship, shared interests, financial exchanges, and relational attributes such as beliefs, knowledge, and credibility. SNA leverages network theory to visualize and analyze social relationships, depicting networks through nodes representing individual actors and edges representing the connections between them. This graph-based approach often reveals complex and intricate structures within the network[61].

Research across numerous scientific disciplines underscores the profound impact of social networks, which operate at scales ranging from familial units to entire nations. These networks play a crucial role in addressing the diverse challenges faced by individuals and organizations, significantly influencing the achievement of specific objectives. At its core, a social network provides a visual representation of specified relationships, such as friendships, existing between nodes [62]. Each node represents an individual, and their connections to other nodes illustrate their social ties. Furthermore, social networks can quantify social capital—the value an individual derives from their position within the network.

It is essential to recognize that social networks encapsulate the intricate web of relationships and influences that connect individuals, rather than merely depicting individual characteristics. These fundamental concepts are often illustrated through social network diagrams, where nodes are shown as points and relationships as lines.

### *C. Indicators for Analyzing Social Networks*

Social networks are characterized by various attributes, which can be categorized into structural, interactive, and functional dimensions. Each category comprises specific indicators chosen based on the research problem and objectives [63].

Structural attributes pertain to the network's configuration, including factors such as size, density, and composition [64].

Interactive characteristics examine the qualities of relationships between members, encompassing parameters like contact frequency, strength, multiplicity, proximity, and duration of relationships[65].

Functional attributes focus on the roles the network plays for its members, such as providing different types of social support[66].

The present research focuses on the network's structural configuration, specifically examining key construction-related indicators, commonly referred to as centrality indicators[67].

1. Degree Centrality:

Degree centrality measures the level of activity or communication a node has with other nodes in the network [68]. This metric provides insights into the proficiency, influence, or experience of network members, identifying which nodes hold greater significance and impact within the network. The degree centrality ($C_d$) of node $v_i$ is the number of edges connected to it, formalized as:

$$C_d(v_i) = \sum_{i=1}^{n} a(v_i, v_j)$$

where $n$ is number of nodes, and $a(v_i, v_j)$ is the edge between nodes $v_i$ and $v_j$.

2. Betweenness Centrality:

Betweenness centrality assesses a node's role in facilitating communication between other nodes, reducing the number of intermediaries needed for information flow. It quantifies how often a node appears on the shortest path between any pair of nodes in the network. Higher betweenness centrality values indicate strategic positioning, as the removal of such a node would disrupt information flow. Denote by $g_{s,t}$ The shortest paths, and let $g_{s,t}$ Be the number of shortest paths passing through some vertex $V$ other than $s, t$ Then this metric is calculated as [62]:

$$C_B(i) = \frac{1}{n^2} \sum_{\forall s,t \in V} \frac{n_{s,t}^{i}}{g_{s,t}}$$

3. Closeness Centrality:

Closeness centrality evaluates a node's accessibility to other nodes within the network. It is computed by taking the inverse of the average shortest path length between the node and all other nodes, with values ranging between 0 and 1. Higher values indicate greater proximity and shorter average distances to other nodes[69, 70]. This metric is formalized as:

$$C_c(v) = \frac{n-1}{\sum_{i=1}^{n} d_{(v,i)}}$$

where $n$ is the number of nodes and $d_{(v,i)}$ is the shortest path between nodes $v$ and $i$.

Centrality measures in social network analysis have been shown to effectively capture the importance and influence of nodes in complex networks [71, 72]. In this context, these measures help identify videos that are well-connected and potentially influential within the viewing network, thus likely to be of interest to users. Specifically, degree centrality captures the direct popularity of videos, closeness centrality identifies videos that are easily accessible to diverse user preferences, and betweenness centrality highlights videos that bridge different viewing communities [73, 74]. This theoretical foundation provides a mathematically sound basis for using network centrality in video recommendation.

In the realm of social network analysis, centrality indicators such as degree, betweenness, and closeness centrality serve as fundamental tools. They provide critical insights into the positions of nodes, their roles in communication, and their influence within the complex fabric of the network. By understanding these indicators, researchers can better analyze the dynamics of social networks and the interdependencies that shape them [62].

### *D. Clustering in Social Networks Analysis*

In real-world scenarios, a cluster often consists of individuals with similar economic, social, or political interests who live near each other [75]. Conversely, virtual clusters form when users connect via social media and interact. For a cluster to form, there must be at least two connections sharing a common interest and commitment to it [76]. A cluster can be described as a group of entities that are closer to each other than to other entities in the dataset [77]. These groups emerge when individuals interact more frequently within the group than with those outside of it (Figure 1). The proximity within a cluster is assessed by examining the similarity or distance between entities. Essentially, a social network cluster is similar to a community [68].

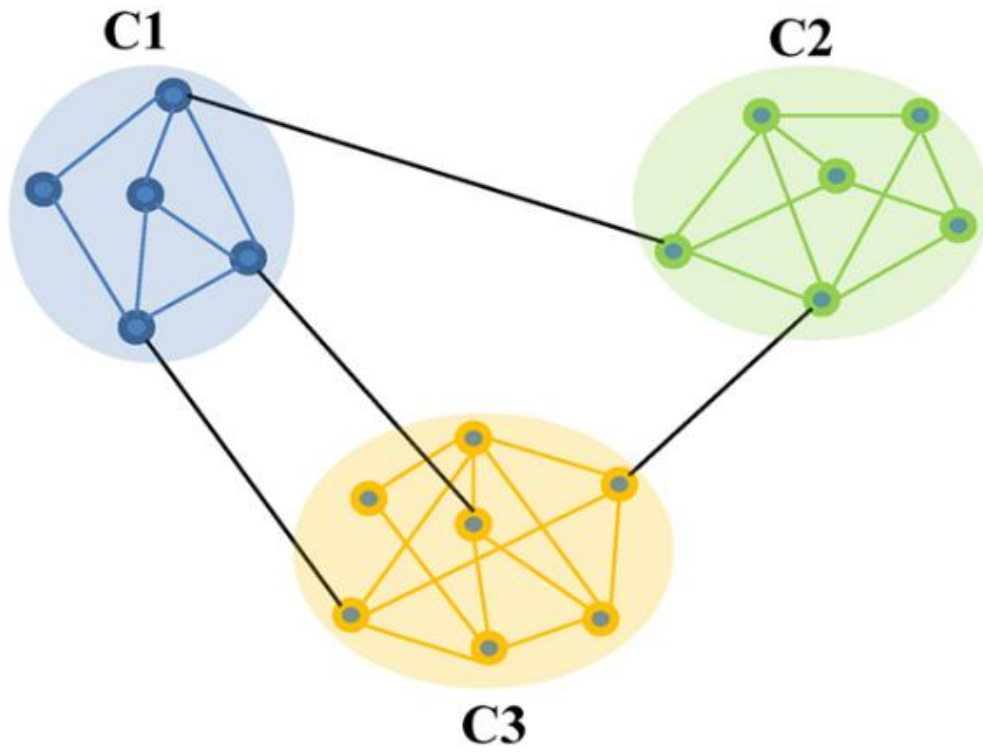

Figure 1: Clusters in social networks[78]

Clustering in network models, as shown in Figure 2, involves identifying clusters as groups of closely associated nodes with stronger connections within the group than with nodes outside it [79]. Clustering helps network analysts understand interactions and cohesive sub-groups within a network. Modularity, as defined by Newman (2006) [80], measures the effectiveness of clustering in social networks. The algorithm for detecting clusters in a weighted network with n nodes starts by treating each node as its cluster. It then searches for a neighboring cluster for each node that maximizes the modularity index when the node is moved. If moving the node increases modularity, it joins the new cluster; otherwise, it remains in its original cluster. This process continues for all nodes until no further changes are possible, reaching a locally optimal point. In the second phase, small clusters merge to form larger ones until the maximum modularity index is achieved. Figure 2 Illustrates how the algorithm initially identifies four clusters and then merges them into two larger clusters to maximize the modularity index [81].

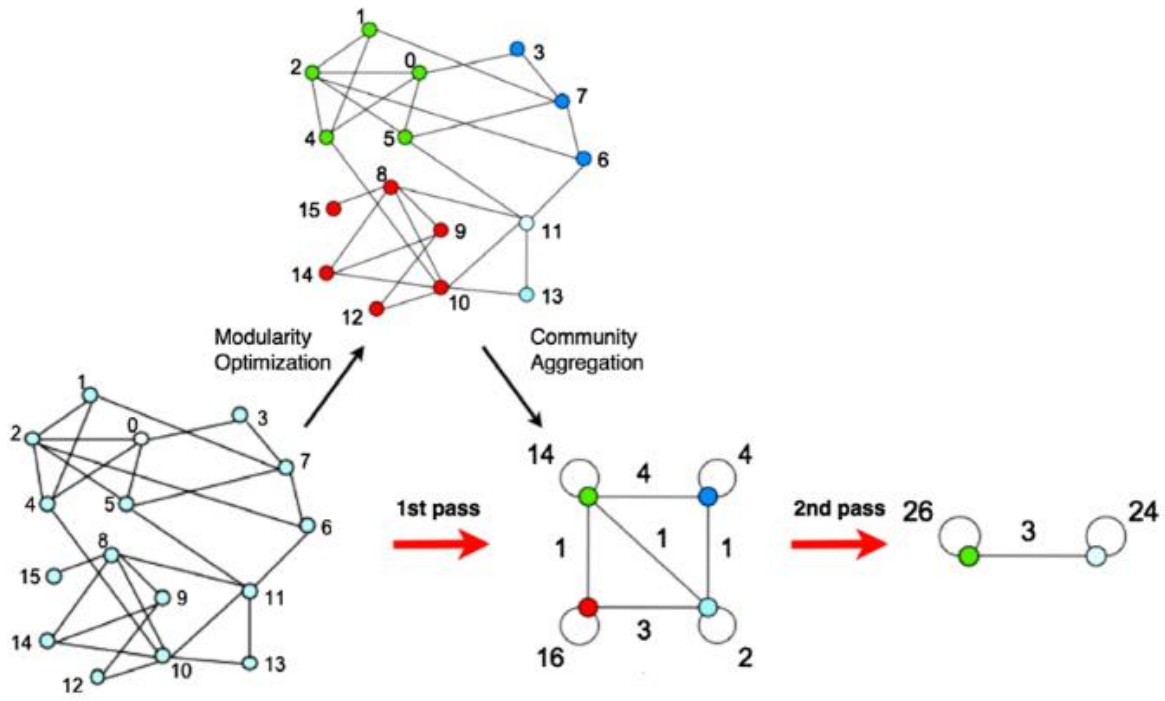

Figure 2: Cluster detection based on increasing modularity[81]

### *E. Background Research Overview*

Innovative approaches in movie recommender systems have been the subject of extensive research and development in recent years. Various studies have explored different methodologies and techniques to enhance the accuracy, efficiency, and scalability of movie recommendation systems. A comparative analysis of these approaches is presented in Table 1.

One prevalent theme in these studies is the utilization of hybrid approaches that combine multiple recommendation strategies to overcome the limitations of individual methods [82]. For instance, a hybrid movie recommendation system was developed, which integrated genetic algorithms and k-means clustering to improve the recommendation quality [83]. This hybrid approach leverages the strengths of both collaborative filtering and content-based filtering to provide more personalized and accurate movie suggestions to users.

Moreover, the incorporation of sentiment analysis has emerged as a promising avenue to enhance movie recommendations. Wang et al. (2018) proposed a sentiment-enhanced hybrid recommender system that leverages big data analytics to improve recommendation efficiency and scalability. By analyzing user sentiments and preferences, this system can offer more tailored movie suggestions that align with individual tastes and preferences. Additionally, sentiment analysis has been used in conjunction with microblogging data to develop movie recommendation systems that take into account user emotions and opinions [84]. This approach enables the system to recommend movies based on the sentiment expressed in user-generated content, leading to more contextually relevant recommendations.

The integration of machine learning algorithms has become central to advancing movie recommendation systems. Machine learning techniques offer the ability to process and analyze massive datasets, detect complex patterns, and make predictions with high accuracy. Among the widely used algorithms are Support Vector Machines (SVM), Genetic Algorithms (GA), and Artificial Neural Networks (ANNs), which have demonstrated remarkable success in tailoring recommendations to individual users' preferences [85-87].

Deep learning has emerged as a transformative force in movie recommendation systems. Convolutional Neural Networks (CNNs) and Recurrent Neural Networks (RNNs), for instance, are applied to analyze both static and sequential data. CNNs are employed to extract visual features from movie posters and frames, while RNNs handle sequential user interaction data to capture dynamic changes in preferences over time. These models excel at identifying complex patterns and generating recommendations that align more closely with user interests [88].

One significant application of deep learning is autoencoders, which are used for dimensionality reduction and feature extraction. In collaborative filtering, for example, autoencoders compress user-item matrices into lower-dimensional representations, enabling systems to process sparse data more effectively. Similarly, variational autoencoders (VAEs) have been explored for generating latent features of users and movies, improving recommendation quality [89, 90].

Reinforcement Learning (RL) introduces an adaptive aspect to recommendation systems by continuously learning and optimizing based on user feedback. In movie recommender systems, RL agents are trained to maximize long-term user satisfaction by balancing exploitation (recommending familiar movies) and exploration (introducing new and diverse content). Deep Q-learning (DQL) and Policy Gradient Methods are commonly employed to refine the recommendation strategies dynamically [91].

Graph-based methods, such as Graph Neural Networks (GNNs), have also been integrated into movie recommendation systems to leverage the relational structure of data. By representing movies, users, and their interactions as a graph, GNNs can capture the complex relationships and dependencies among entities. For instance, a GNN can identify clusters of users with similar tastes and recommend movies that are popular within these communities [88]. This approach not only enhances recommendation accuracy but also addresses challenges like the cold-start problem for new users and movies.

Furthermore, the integration of machine learning algorithms has been a key focus in advancing movie recommender systems. Researchers have explored the application of machine learning techniques such as support vector machines, genetic algorithms, and artificial neural networks to enhance the recommendation process. These methods are particularly useful in analyzing vast datasets to identify patterns and generate personalized recommendations for users [88]. For instance, genetic algorithms have been used to optimize the parameters of support vector machines, resulting in improved accuracy for classification tasks, including those in recommender systems [92].

Deep learning methods have also been extensively investigated for improving the accuracy and performance of movie recommendation systems. Models like convolutional neural networks (CNNs) and recurrent neural networks (RNNs) have been employed to extract intricate features from movie data, such as metadata, visual content, and sequential viewing patterns. These models enable more nuanced and precise recommendations that align with user preferences [93].

In the realm of content-based recommendation systems, researchers have leveraged diverse features, including metadata, visual, and audio attributes, to improve recommendation quality. For example, integrating visual features like cinematography or poster design allows systems to cater to users' aesthetic preferences. Similarly, audio features have been analyzed to tailor recommendations for users interested in specific soundtracks or soundscapes [94].

Additionally, the exploration of aesthetic features in visual content has been proposed as a means to improve movie recommendations. By analyzing attributes such as cinematography and visual composition, recommender systems can identify visually appealing movies for users with preferences centered on film artistry [88].

Moreover, the development of real-time and personalized movie recommendation systems has been a focal point in recent research endeavors. A personalized real-time movie recommendation system was designed that utilizes data clustering and computational intelligence techniques to enhance recommendation accuracy [95]. By incorporating algorithms such as K-means clustering and cuckoo search optimization, this system can adapt to user preferences in real-time, providing up-to-date and relevant movie suggestions. Additionally, the integration of chatbot technology has been explored to create interactive and user-friendly movie recommender systems. Chatbots enable users to receive recommendations through natural language interactions, simplifying the recommendation process and enhancing user engagement.

Presently, this research endeavors to introduce an innovative content-based filtering method, rooted in user video viewing data and the utilization of social network analysis indicators. The unique aspect of this method lies in its minimal data requirement—specifically, the duration of video viewing by the user. The subsequent sections delineate the operational steps of this algorithm.

TABLE 1: A COMPARATIVE ANALYSIS OF APPROACHES

| *Approach* | *Methodology* | *Advantages* | *References* |
|---|---|---|---|
| **Hybrid Approaches** | Combines multiple recommendation strategies (e.g., genetic algorithms and k-means clustering). | Enhances recommendation quality by leveraging strengths of collaborative and content-based filtering. | [82, 83] |
| **Sentiment Analysis Integration** | Utilizes sentiment analysis with big data analytics and microblogging data. | Provides more tailored and contextually relevant movie suggestions based on user emotions and opinions. | [84] |
| **Machine Learning Algorithms** | Applies algorithms like SVM, GA, and ANNs to process large datasets and detect complex patterns. | Improves personalization and accuracy of recommendations by analyzing vast amounts of user data. | [85-87, 92] |
| **Deep Learning Methods** | Employs CNNs and RNNs to extract features from visual content and sequential user interactions. | Identifies intricate patterns for more precise and nuanced recommendations aligned with user interests. | [88, 93] |
| **Autoencoders** | Uses autoencoders and VAEs for dimensionality reduction and feature extraction in collaborative filtering. | Enhances recommendation quality by effectively processing sparse data and generating latent features. | [89, 90] |
| **Reinforcement Learning (RL)** | Implements RL agents like DQL and Policy Gradient Methods to balance exploitation and exploration. | Continuously optimizes recommendations based on user feedback, improving long-term user satisfaction. | [91] |
| **Graph-Based Methods** | Integrates GNNs to represent movies, users, and interactions as graphs for relational analysis. | Captures complex relationships, improves accuracy, and addresses cold-start problems for new users and movies. | [88] |
| **Content-Based Filtering** | Leverages metadata, visual, and audio attributes to cater to specific user preferences. | Provides personalized recommendations based on detailed content features like cinematography and soundtrack preferences. | [94] |
| **Real-Time Personalized Systems** | Utilizes data clustering and computational intelligence techniques for up-to-date recommendations. | Adapts to user preferences in real-time, enhancing accuracy and relevance of movie suggestions. | [95] |
| **Chatbot Integration** | Incorporates chatbot technology for interactive and user-friendly recommendation processes. | Simplifies the recommendation process through natural language interactions, increasing user engagement. | [95] |

## III. Research Methodology

The proposed methodology for creating a movie recommendation system utilizes user viewing history and incorporates social network analysis indicators. The steps involved in this approach are as follows:

**Step 1: Create a User-Viewing Percentage Matrix**

Using data on user movie-watching habits, construct a matrix showing the percentage of each movie watched by individual users (Table 2).

TABLE 2: USER VIEWING PERCENTAGE MATRIX

| *Movie ID* | *User ID* | *Duration of watching the video by the user (seconds)* | *Total video duration (seconds)* | *User viewing percentage* |
|---|---|---|---|---|
| F1 | U1 | X1 | Y1 | X1/Y1 |
| F2 | U2 | X2 | Y2 | X2/Y2 |
| F1 | U3 | X3 | Y1 | X3/Y1 |
| F3 | U4 | X4 | Y3 | X4/Y3 |

**Step 2: Calculate Movie Similarity Matrix**

Using the data from Step 1, calculate the similarity between two movies from each user's perspective (Dual Similarity, DS) using equation (1):

$$DS_{ij} = \frac{2n_{ij}}{n_i + n_j} \qquad (1)$$

where $n_i$ is the proportion of movie i watched by user n, $n_j$ is the proportion of movie j watched by user n, and $n_{ij}$ is the simultaneous viewership of movies i and j by user n, calculated using equation (2):

$$n_i \cap n_j = \min\,(n_i\,, n_j\,) \qquad (2)$$

The resulting matrix is shown in Table 3.

TABLE 3: MOVIE SIMILARITY MATRIX

| **i** | **j** | **U1** | **U2** | **U3** | **…** |
|---|---|---|---|---|---|
| *F1* | *F2* | $DS_{F1F2}$ | $DS_{F1F2}$ | $DS_{F1F2}$ | … |
| *F1* | *F3* | $DS_{F1F3}$ | $DS_{F1F3}$ | $DS_{F1F3}$ | … |
| *F2* | *F3* | $DS_{F2F3}$ | $DS_{F2F3}$ | $DS_{F2F3}$ | … |

**Step 3: Construct Average Similarity Matrix**

Using the DS values from Step 2, calculate the average similarity (AS) between two movies using equation (3):

$$AS_{ij} = \frac{\sum_{n=1}^{p} DS_{ij}(n)}{p} \qquad (3)$$

where $p$ is the number of users. The resulting matrix is shown in Table 4.

TABLE 4: AVERAGE SIMILARITY MATRIX

| *Films* | *F1* | *F2* | *F3* | … |
|---|---|---|---|---|
| *F1* | *1* | $AS_{F1F2}$ | $AS_{F1F3}$ | … |
| *F2* | $AS_{F2F1}$ | *1* | $AS_{F2F3}$ | … |
| *F3* | $AS_{F3F1}$ | $AS_{F3F2}$ | *1* | … |
| ….. | … | … | … | *1* |

**Step 4: Create a Movie Relationship Graph**

Using the AS values from Step 3, create a graph where movies are nodes and AS values define the edges. The edge thickness corresponds to AS values.

**Step 5: Compute Centrality Measures**

Using the graph from Step 4, compute three centrality measures for each movie: degree centrality ($D_C$), closeness centrality ($C_C$), and betweenness centrality ($B_C$). Combine these into an average centrality (AC) using equation (4) and the result of this stage is a matrix resembling the format of Table 5:

$$AC(i) = \frac{D_C^W(i)+C_C^W(i)+B_C^W(i)}{3} \qquad (4)$$

TABLE 5: CENTRALITY INDICES FOR MOVIES

| *Films ID* | *Degree Centrality* $D_C$ | *Closeness Centrality* $C_C$ | *Betweenness Centrality* $B_C$ | *Average Centrality* $AC$ |
|---|---|---|---|---|
| *F1* | *0.0536* | *0.8229* | *0.5622* | *0.4796* |
| *F2* | *0.0345* | *0.6475* | *0.1204* | *0.2675* |
| *F3* | *0.0497* | *0.8495* | *0.5207* | *0.4733* |
| …. | *0.0438* | *0.7745* | *0.426* | *0.4148* |

**Step 6: Movie Clustering**

Cluster movies using modularity analysis (Newman, 2006):

$$Q = \sum_r (e_{rr} - a_r^2) \qquad (5)$$

where $e_{rr}$ is the count of connections within a cluster, and $a_r^2$ is the count of connections involving at least one node in the cluster. Higher Q values indicate stronger community structures.

The modularity resolution parameter γ was selected based on an extensive review of existing literature and empirical analysis. Fortunato and Barthélemy (2007) demonstrated that $\gamma = 1$provides a balanced approach between detecting small, cohesive communities and larger, more inclusive ones. To ensure the robustness of our clustering results, this research conducted a comprehensive sensitivity analysis by varying $\gamma$ from 0.5 to 2.0 in increments of 0.1. This analysis revealed that $\gamma = 1$ consistently yielded the most stable and interpretable clusters across different metrics, including silhouette scores and intra-cluster similarity measures. Additionally, clusters formed at $\gamma = 1$ significantly improved recommendation quality, as evidenced by higher precision and recall rates in our evaluation metrics compared to other $\gamma$ values. Therefore, $\gamma = 1$ was adopted as the optimal resolution parameter for our dataset, ensuring that the clustering effectively captures meaningful groupings that enhance the recommendation system's performance.

**Step 7: Create Preference and Non-Preference Matrices**

Classify movies as preferred if the viewing percentage is over 50% and as non-preferred otherwise (Table 6).

TABLE 6: USER PREFERENCES AND NON-PREFERENCES

| *User* | *List of Non-Preferences* | | | | *Favorites List* | |
|---|---|---|---|---|---|---|
| *U1* | *Film10* | *Film7* | *Film2* | *Film43* | *Film4* | *Film1* |
| *U2* | *Film4* | *Film13* | *Film5* | | *Film6* | *Film2* |
| …. | ….. | ….. | ….. | | …. | ….. |

**Step 8: Compute the Ego-Focused Centrality Index**

For network analysis, two distinct methodologies are employed: socio-centric and ego-focused approaches. These approaches are depicted in Figure 3.

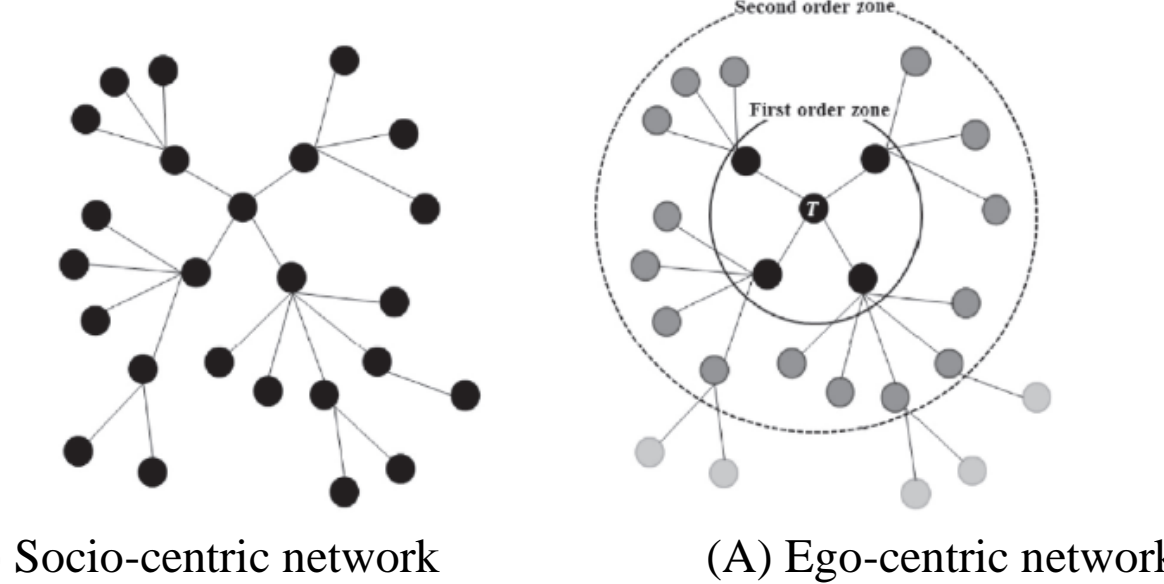

Figure 3: An example of ego-centric and socio-centric networks

The indices computed in the fifth step correspond to socio-centric metrics. A higher value of the average centrality outlined in equation (4) indicates that a particular movie enjoys more popularity than other items. This methodology is applied for providing recommendations to users without a preexisting viewing history. However, for personalized recommendations, the information linked to each user's preferences needs to play a significant role in the suggested offerings.

When a movie captures a user's interest, it becomes the focal point (ego node), and the recommendation system seeks out other items akin to this selection. The potential candidate items are selected from among the members of the user's preferences co-cluster.

Referring to Figure 4, within this ego-centric network, node $T$ Holds the position of the central node, denoting a user's preference. The initial layer surrounding this node is composed of direct connections to it. The subsequent layer is connected to the node $T$ via at least one intermediary node, thus establishing the relational structure between node $T$ and all other nodes within the network.

Based on these explanations, the ego-centric centrality, designated as $C_{EF}$ Is delineated as follows.

$$C_{EF}(i.T) = \frac{AC(i)}{d(i.T)} \qquad (6)$$

In this context, the value of $AC(i)$ will be derived using equation (4), and the distance between node $i$ and node $T$ Will be assessed based on the count of links that connect them. This distance represents the number of links separating the two nodes.

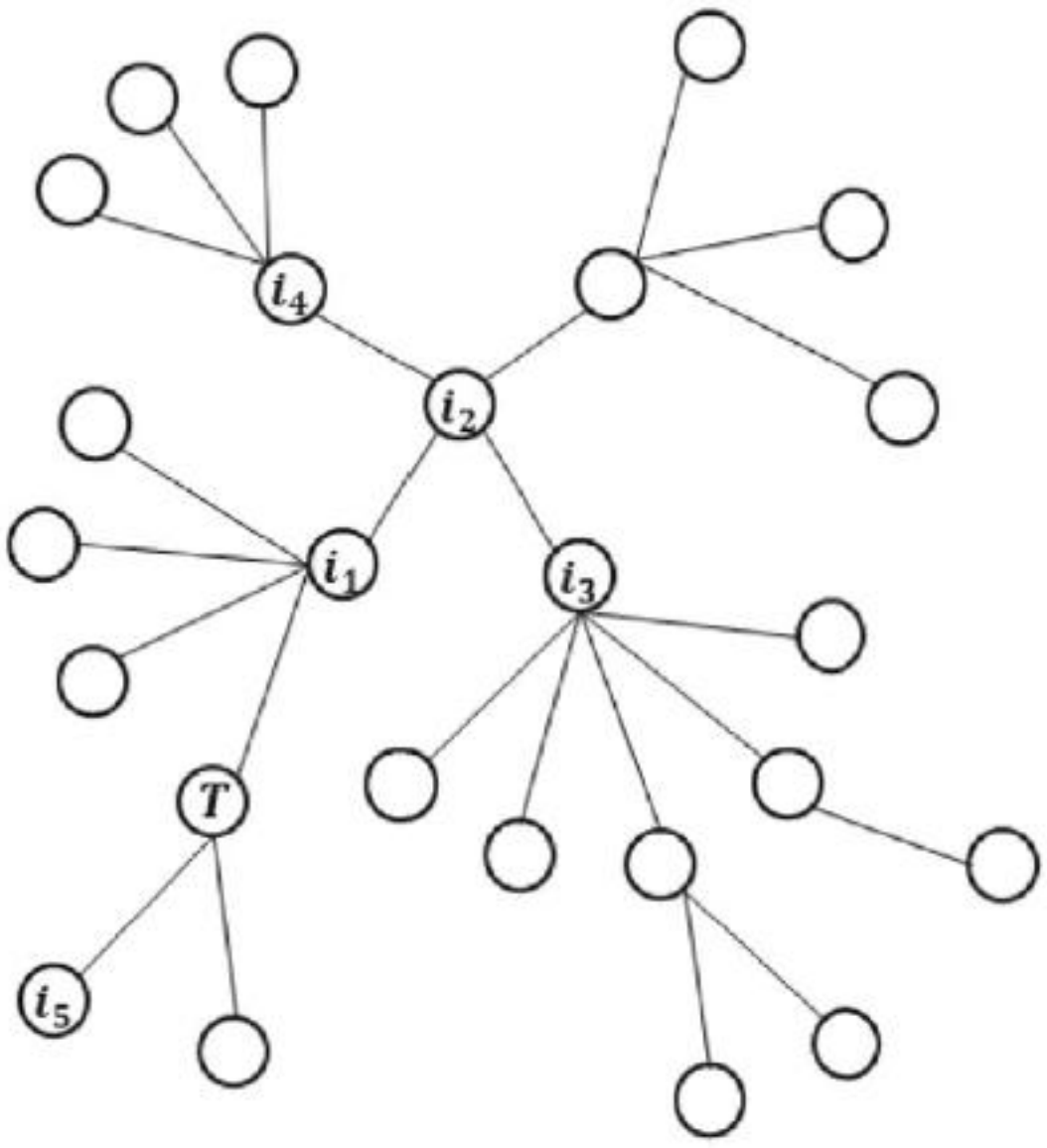

Figure 4: An ego-centric network whose central node is T.

As an example, in Table 7, despite node $i_1$ having a lower average centrality value compared to the node $i_3$, its ego-centric centrality has increased due to its closer distance from the central node $i_1$.

TABLE 7: EGO-CENTRIC CENTRALITY CALCULATION

| Node | Ego-centric Centrality | Node distance | Average Centrality $AC$ | Betweenness Centrality $B_C$ | Closeness Centrality $C_C$ | Degree Centrality $D_C$ |
|---|---|---|---|---|---|---|
| i1 | 0.739 | 1 | 0.739 | 0.567 | 0.817 | 0.833 |
| i2 | 0.445 | 2 | 0.889 | 1 | 1 | 0.667 |
| i3 | 0.288 | 3 | 0.865 | 0.719 | 0.875 | 1 |
| i4 | 0.175 | 3 | 0.524 | 0.163 | 0.742 | 0.667 |
| i5 | 0.216 | 1 | 0.216 | 0 | 0.48 | 0.167 |

**Step 9: Create the Ego-Centric Recommendation Matrix**

Construct the ego-centric centrality index matrix using user preferences and non-preferences. Table 8 illustrates the outcomes.

TABLE 8: EGO-CENTRIC RECOMMENDATION SCORES

| User ID | | | | | | |
|---|---|---|---|---|---|---|
| *Candidate movie* | *List of preferences* | | | *List of non-preferences* | | $RS_{EF}$ |
| | *F10* | *F11* | *F9* | *F3* | *F13* | |
| *F1* | $C_{EF}(F13, F1)$ | $C_{EF}(F3, F1)$ | $C_{EF}(F9, F1)$ | $C_{EF}(F11, F1)$ | $C_{EF}(F10, F1)$ | *0* |
| *F2* | $C_{EF}(F13, F2)$ | $C_{EF}(F3, F2)$ | $C_{EF}(F9, F2)$ | $C_{EF}(F11, F2)$ | $C_{EF}(F10, F2)$ | *0.1337* |
| *F7* | $C_{EF}(F113, F7)$ | $C_{EF}(F3, F7)$ | $C_{EF}(F9, F7)$ | $C_{EF}(F11, F7)$ | $C_{EF}(F10, F7)$ | *0.1442-* |
| …. | …. | …. | …. | …. | …. | …. |

The recommendation score $RS_{EF}$ Calculated using:

$$RS_{EF}(i) = (\sum_{j=1} C_{EF-P}(i.T_j)) - \sum_{k=1} C_{EF-NP}(i.T_k)) \qquad (7)$$

Where $C_{EF-P}$ and $C_{EF}$ Are the ego-centric centrality indices for preferences and non-preferences, respectively?

The sequence of steps for proposing items in this research is illustrated in Figure 5.

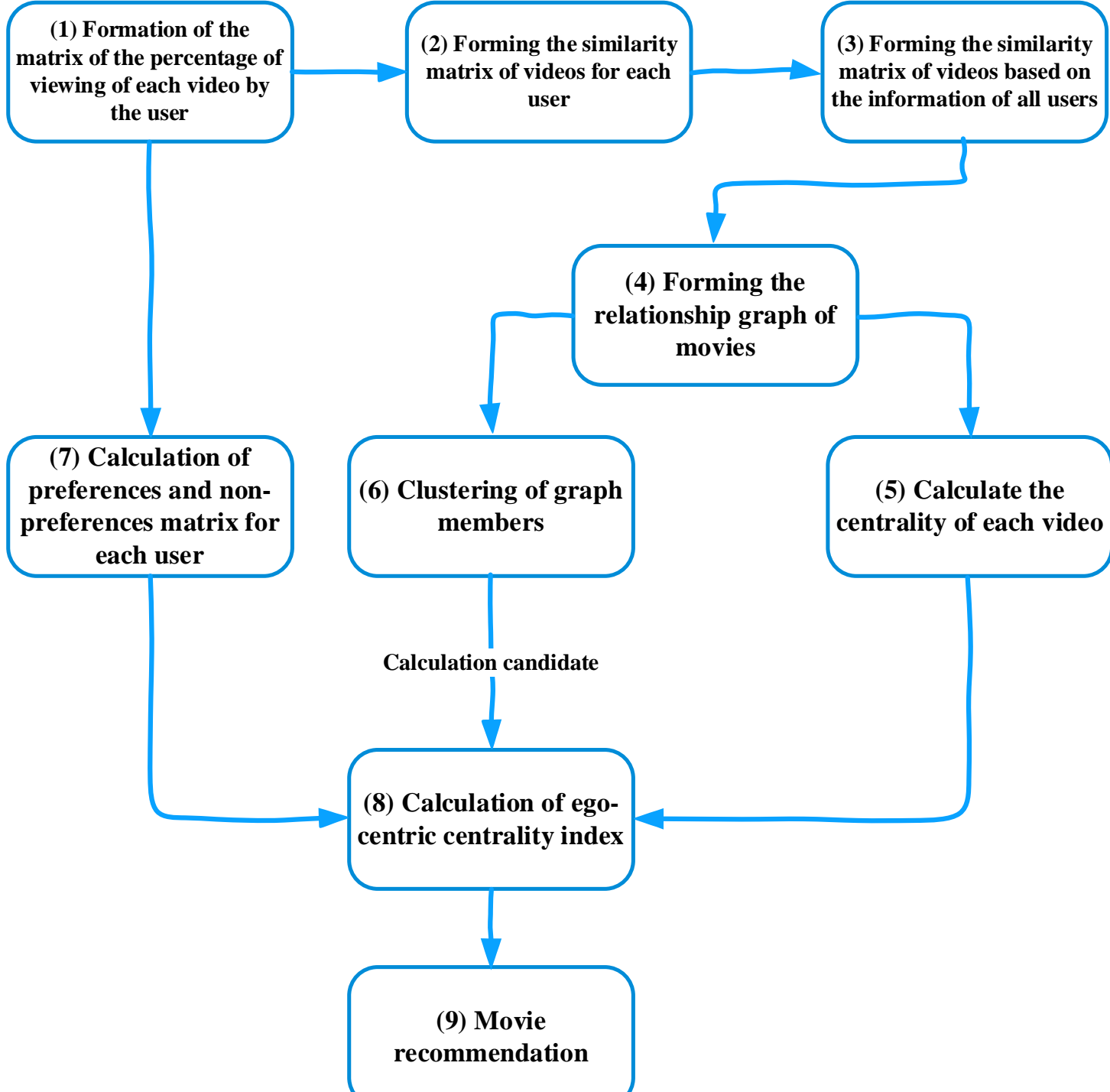

Figure 5: Steps of research model for movie recommendation

### B. Data used

The dataset consists of user viewing details from an Asian video-on-demand platform over 10 months, from January 1, 2018, to September 30, 2018. It includes 80 movie titles and 328 users. The platform was established in 2013 and offers both streaming and downloadable content, focusing primarily on documentaries.

It should be noted that this dataset comes from a specialized documentary-focused VOD platform, making it particularly suitable for proof-of-concept validation of the proposed methodology. While the dataset size is modest compared to general-purpose platforms, it provides a controlled environment for testing the effectiveness of network-based recommendation approaches in a specialized content domain.

## IV. RESULTS

### A. User Video Viewing Percentage Matrix

The first step of this study involves creating a table that shows the percentage of each video watched by each user. This percentage is calculated by dividing the time a user spends watching a video by the video's total duration. Table 9 Presents a portion of this data.

TABLE 9: USER VIDEO VIEWING DATA

| *User viewing percentage* | *Total video duration (seconds)* | *Duration of watching the video by the user (seconds)* | *User number* | *Movie number* |
|---|---|---|---|---|
| 0.98 | 2458 | 2400 | 59635 | 1401 |
| 0.22 | 2469 | 540 | 44530 | 6352 |
| 0.68 | 2469 | 1680 | 71326 | 6352 |
| 0.99 | 2533 | 2520 | 58197 | 236 |

| 0.4 | 2472 | 1020 | 59243 | 53 |
|---|---|---|---|---|
| 0.82 | 2572 | 2100 | 884 | 53 |
| 0.96 | 2572 | 2460 | 59100 | 53 |
| ... | ... | ... | ... | ... |

### B. User-Based Video Similarity Matrix

The next step involves creating a similarity matrix for videos based on user views using equation (1). If both the numerator and denominator in the equation are zero, the value -1 is used to indicate that the user has not watched either movie. A value of 0 means the user watched one but not the other, implying no similarity. Table 10 Shows part of this 80x324 matrix.

TABLE 10: USER-BASED VIDEO SIMILARITY MATRIX

| | | *user 73094* | *user 70324* | .... | *user 5383* | *user 4486* |
|---|---|---|---|---|---|---|
| movie 61 | **movie 51** | -1 | -1 | ... | -1 | 0 |
| movie 5490 | **movie 61** | -1 | -1 | ... | -1 | 0.82 |
| movie 5591 | **movie 61** | -1 | -1 | ... | -1 | 0.99 |
| ... | **....** | ... | ... | ... | ... | ... |
| movie 53 | **movie 236** | -1 | -1 | ... | -1 | -1 |
| movie 590 | **movie 590** | -1 | -1 | ... | 1 | 1 |
| movie 6518 | **movie 590** | -1 | 0 | ... | 0.71 | 0.53 |

### C. Square Matrix of Movie Similarity

In the third phase, this research calculates the average connection strength (similarity) between each pair of movies using equation (2).

TABLE 11: SQUARE MATRIX OF MOVIE SIMILARITY

| | *Movie 6709* | *Movie 6518* | ... | *Movie 61* | *Movie 51* |
|---|---|---|---|---|---|
| Movie 51 | 0.0901 | 0.0381 | ... | 0.0194 | 1 |
| Movie 61 | 0.0285 | 0.0439 | ... | 1 | 0.0194 |
| ... | ... | ... | ... | ... | ... |
| Movie 6518 | 0.0537 | 1 | ... | 0.0439 | 0.0381 |
| Movie 6709 | 1 | 0.0537 | ... | 0.0285 | 0.0901 |

### D. Graph of Movie Relationships

Using the data from Table 11, this research constructed a graph to illustrate the relationships among movies. The graph is presented in Figure 6.

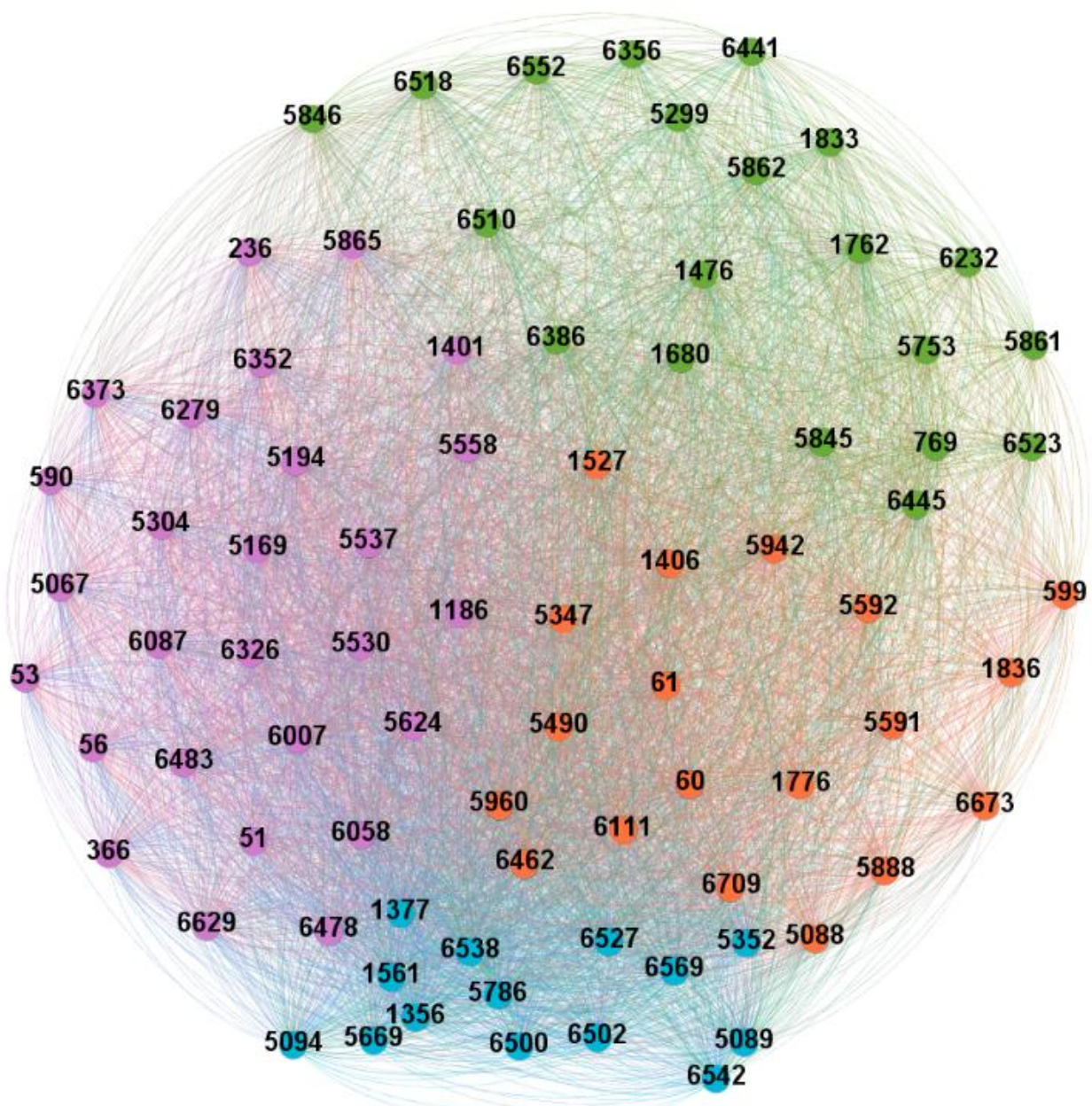

Figure 6: Movie Relationship Graph Based on Centrality

### E. *Calculation of Movie Centrality Indexes*

Table 12TShows12hows the calculated centrality indices for each movie, including Degree Centrality, Closeness Centrality, Betweenness Centrality, and Average Centrality.

TABLE 12: CENTRALITY INDICES OF MOVIES

| *Movie Number* | *Degree Centrality* | *Closeness Centrality* | *Betweenness Centrality* | *Average Centrality* |
|---|---|---|---|---|
| 51 | 0.0536 | 0.8229 | 0.5622 | 0.4796 |
| 53 | 0.0345 | 0.6475 | 0.1204 | 0.2675 |
| 56 | 0.0497 | 0.8495 | 0.5207 | 0.4733 |
| 60 | 0.0438 | 0.7745 | 0.426 | 0.4148 |
| 61 | 0.043 | 0.7383 | 0.2711 | 0.3508 |
| ... | ... | ... | ... | ... |

### F. *Video Clustering*

Using modularity analysis, movies were clustered.

Table 13 and the colors in Figure 6 Show the clustering results. The movies in the most preferred cluster are selected as recommendation candidates.

TABLE 13: CLUSTERING OF MOVIES

| *Cluster 1* | | *Cluster 2* | | *Cluster 3* | | *Cluster 4* | |
|---|---|---|---|---|---|---|---|
| movie 236 | movie 366 | movie 51 | movie 60 | movie 53 | movie 61 | movie 56 | movie 1680 |
| movie 1356 | movie 1401 | movie 590 | movie 599 | movie 769 | movie 1476 | movie 1776 | movie 1833 |
| | | movie 1186 | movie 1377 | movie 1762 | movie 1836 | movie 5094 | movie 5169 |
| movie 1527 | movie 5194 | movie 1406 | movie 1561 | movie 5299 | movie 5490 | movie 5669 | movie 5861 |
| movie 5304 | movie 5865 | movie 5067 | movie 5088 | movie 5591 | movie 5592 | movie 6279 | movie 6462 |
| movie 6232 | | movie 5089 | movie 5347 | movie 5753 | movie 5845 | movie 6500 | movie 6502 |
| | | movie 5352 | movie 5530 | movie 5846 | movie 5862 | movie 6527 | movie 6552 |
| | | movie 5537 | movie 5558 | movie 6356 | movie 6373 | movie 6569 | |
| | | movie 5624 | movie 5786 | movie 6386 | movie 6441 | | |
| | | movie 5888 | movie 5942 | movie 6445 | movie 6510 | | |
| | | movie 5960 | movie 6007 | movie 6518 | movie 6523 | | |
| | | movie 6058 | movie 6087 | movie 6538 | movie 6542 | | |
| | | movie 6111 | movie 6326 | movie 6629 | movie 6673 | | |
| | | movie 6352 | movie 6478 | | | | |
| | | movie 6483 | movie 6709 | | | | |

### G. User Preferences and Non-Preferences Matrix

This research calculated user preferences and non-preferences based on their movie-watching data. Table 14 shows part of these results.

TABLE 14: USER PREFERENCES AND NON-PREFERENCES

| *Users* | *Preferences list (favorite movies)* | | | *List of non-preferences (uninterested films)* | | |
|---|---|---|---|---|---|---|
| user 4486 | movie 6523 | movie 5591 | movie 590 | movie 6709 | movie 5088 | movie 236 |
| user 5383 | | movie 1377 | movie W590 | | movie 6552 | movie 6538 |
| ... | ... | ... | ... | ... | ... | ... |
| user 70324 | | movie 6356 | movie 5304 | movie 6569 | movie 6500 | movie 836 |
| user 73094 | | movie 6352 | movie 5537 | movie 5592 | movie 5089 | movie 236 |

### H. Calculation of Ego-Centered Centrality Index

Using equation (6), the value of $C_{EF}$ was calculated for the preferences and non-preferences of user 5383. Table 15 to Table 18 present these calculations.

TABLE 15: $C_{EF}$ CALCULATION WITH EGO-CENTRIC MOVIE 590 FOR USER 5383

| *Movie number* | *Ego-centric* | *Link distance to movie 590* | *Average centrality* |
|---|---|---|---|
| 51 | 0.4796 | 1 | 0.4796 |
| 53 | 0.1337 | 2 | 0.2675 |
| ... | ... | ... | ... |
| 6673 | 0.1442 | 2 | 0.2885 |
| 6709 | 0.2743 | 1 | 0.2743 |

TABLE 16: $C_{EF}$ CALCULATION WITH EGO-CENTRIC MOVIE 1377 FOR USER 5383

| *Movie Num.* | *Average Centrality* | *Link distance to movie 1377* | *Ego-centric Centrality* |
|---|---|---|---|
| 51 | 0.4796 | 1 | 0.4796 |
| 53 | 0.2675 | 1 | 0.2675 |
| ... | ... | ... | ... |
| 6673 | 0.2885 | 1 | 0.2885 |
| 6709 | 0.2743 | 1 | 0.2743 |

TABLE 17: $C_{EF}$ CALCULATION WITH EGO-CENTRIC MOVIE 6538 FOR USER 5383

| *Movie Num.* | *Average Centrality* | *Link distance to movie 6538* | *Ego-centric Centrality* |
|---|---|---|---|
| 51 | 0.4796 | 2 | 0.2398 |
| 53 | 0.2675 | 1 | 0.2675 |
| ... | ... | ... | ... |
| 6673 | 0.2885 | 2 | 0.1442 |
| 6709 | 0.2743 | 2 | 0.1371 |

TABLE 18: $C_{EF}$ CALCULATION WITH EGO-CENTRIC MOVIE 6552 FOR USER 5383

| *Movie Num.* | *Average Centrality* | *Link distance to movie 6552* | *Ego-centric Centrality* |
|---|---|---|---|
| 51 | 0.4796 | 1 | 0.4796 |
| 53 | 0.2675 | 2 | 0.1337 |
| ... | ... | ... | ... |
| 6673 | 0.2885 | 1 | 0.2885 |
| 6709 | 0.2743 | 1 | 0.2743 |

### *I. Table of Centrality Indexes of Preference and Non-Preference Values*

Table 19 shows the calculated centrality index values for the preference and non-preference lists of users.

TABLE 19: CENTRALITY INDEX VALUES FOR USER PREFERENCES AND NON-PREFERENCES

| | *List of preferences* | | *List of non-preferences* | | *RSEF* |
|---|---|---|---|---|---|
| | movie 1377 | movie 590 | movie 6552 | movie 6538 | |
| 51 | 0.4796 | 0.4796 | 0.4796 | 0.4796 | 0 |
| 53 | 0.2675 | 0.1337 | 0.1337 | 0.1337 | 0.1337 |
| ... | ... | ... | ... | ... | ... |
| 6673 | 0.2885 | 0.1442 | 0.2885 | 0.2885 | 0.1442- |
| 6709 | 0.2743 | 0.2743 | 0.2743 | 0.2743 | 0 |

According to the calculations made for this example, movie number 53 is preferable to the other 3 items in offering to this user, and movie number 6673 should not be offered to this user.

The improvements in click-through rate and view completion rates highlight the effectiveness of clustering closely related videos and leveraging centrality measures for identifying content that aligns with user preferences. These results suggest that modularity-based clustering captures meaningful video relationships, while ego-centric ranking enhances user engagement by focusing on familiar content proximities.

### *J. Model evaluation*

Evaluating and validating the proposed research model is a critical component of this study, as the model's effectiveness directly impacts the credibility of the underlying research assumptions. The primary metric utilized for this evaluation is the $RS_{EF}$ Value, which quantifies the model's ability to accurately identify user preferences based on viewership data. Specifically, for each user, two movies with the highest viewership rates and two movies with the lowest viewership rates were designated as the user's preferred and non-preferred items, respectively. A positive $RS_{EF}$ Value is expected for preferred movies, while a negative value should correspond to non-preferred movies. Accurate identification of these preferences by the model would result in the expected $RS_{EF}$ values, thereby confirming the model's accuracy. Conversely, deviations from these expected values would indicate inaccuracies in the model's predictions.

Additionally, to provide a more comprehensive evaluation of the recommendation system, this research incorporated diversity and novelty measures alongside accuracy-based metrics.

**Diversity** is measured using the **Intra-List Diversity (ILD)** metric, which calculates the average dissimilarity between all pairs of recommended items for a user. A higher ILD value indicates greater diversity in the recommendations [91].

$$ILD = \frac{2}{N(N-1)} \sum_{i=1}^{N} \sum_{j=i+1}^{N} (1 - \mathrm{sim}(m_i, m_j))$$

where $N$ is the number of recommended movies, and $\mathrm{sim}(m_i, m_j)$represents the similarity between movies $m_i$ and $m_j$.

**Novelty** is assessed using the **Average Popularity (AP)** metric, which evaluates how novel the recommended movies are by measuring the average popularity of the recommended items. Lower AP values signify higher novelty, indicating that the recommendations include less popular (and potentially more novel) movies [88].

$$AP = \frac{1}{N} \sum_{i=1}^{N} \mathrm{popularity}\,(m_i)$$

where $N$ is the number of recommended movies, and $\mathrm{popularity}(m_i)$ denotes the popularity score of movie $m_i$.

The calculation of the Intra-List Diversity (ILD) and Average Popularity (AP) metrics was implemented using Python, leveraging libraries such as NumPy and pandas for numerical operations and data manipulation. For ILD, cosine similarity between recommended movies was calculated using the scikit-learn library's *cosine_similarity* function. The ILD was derived by averaging the dissimilarities across all pairs of recommended movies. For AP, the popularity scores of recommended movies were aggregated based on their frequency in the dataset.

The inclusion of these metrics allows for a balanced evaluation, ensuring that the recommendation system not only predicts user preferences accurately but also provides a diverse and novel set of recommendations that enhance user experience [90].

To demonstrate the application of the $RS_{EF}$ metric, detailed calculations for a specific test user (user 62277) under different training data scenarios are presented in Table 20 and Table 21.

TABLE 20: CALCULATION OF $RS_{EF}$ FOR TEST USER 62277 IN MODE 35 TRAINING DATA

| | *List of preferences* | | *List of non-preferences* | | *RSEF* |
|---|---|---|---|---|---|
| | movie 5089 | movie 6007 | movie 5094 | movie 1561 | |
| movie 6007 | 0.3081 | 0.3081 | 0.1541 | 0.3081 | 0.1541 |
| movie 5089 | 0.1933 | 0.1933 | 0.0967 | 0.1933 | 0.0967 |
| movie 1561 | 0.1788 | 0.1788 | 0.0894 | 0.1788 | 0.0894 |
| movie 5094 | 0 | 0 | | | 0 |

In the scenario with 35 training instances, the model accurately assigned positive $RS_{EF}$ values to the preferred movies (Movies 5089 and 6007). However, it erroneously assigned a positive $RS_{EF}$to Movie 1561, a non-preferred movie, indicating a misclassification.

Table 21: Calculation of $RS_{EF}$ for test user 62277 in mode 70 training data

| | *List of preferences* | | *List of non-preferences* | | *RSEF* |
|---|---|---|---|---|---|
| | movie 5089 | movie 6007 | movie 5094 | movie 1561 | |
| movie 6007 | 0.195 | 0.195 | 0.098 | 0.195 | 0.098 |
| movie 5089 | 0.1666 | 0.1666 | 0.1666 | 0.1666 | 0 |
| movie 1561 | 0.2016 | 0.2016 | 0.1008 | 0.2016 | 0.1008 |
| movie 5094 | 0.2121 | 0.1061 | 0.2121 | 0.1061 | 0 |

Similarly, with 70 training instances, the model correctly identified the preferred movies but again incorrectly classified Movie 1561 as preferred, highlighting a consistent area for improvement.

To evaluate the model's performance across different user sample sizes, three distinct groups comprising 50, 100, and 200 individuals were randomly selected from a total pool of 328 users. Each dataset was partitioned into 70% for training and 30% for testing. The aggregated $RS_{EF}$ values for all test users across varying scenarios—15, 30, and 60 test users paired with 35, 70, and 140 training users—were compared against the expected scoring list. A value of one was assigned when the $RS_{EF}$ value aligned with the expected preferences, zero for mismatches, and negative one for disparities. These comparisons facilitate the assessment of the research model's accuracy relative to conventional prediction methods, as summarized in Figure 7.

Additionally, this research evaluated the diversity and novelty of the recommendations using Intra-List Diversity (ILD) and Average Popularity (AP) metrics. The ILD metric assesses the variety within the recommended lists, while the AP metric measures novelty by examining the average popularity of the recommended movies.

To further substantiate the model's performance, paired t-tests were conducted comparing the proposed method against two established baseline algorithms: Matrix Factorization and Neural Collaborative Filtering.

Matrix Factorization is a collaborative filtering technique that decomposes the user-item interaction matrix into lower-dimensional latent factors, effectively capturing the underlying structure of user preferences and item characteristics [96]. In this study, Matrix Factorization was implemented using the Surprise Python library [97]. The algorithm was trained on 70% of the dataset and tested on the remaining 30%, resulting in performance metrics of Precision: 0.301, Recall: 0.138, NDCG: 0.245, and Runtime: 1.8 seconds, as shown in Table 3.

Neural Collaborative Filtering (NCF) leverages deep learning techniques to model complex user-item interactions by capturing non-linear relationships through neural networks [98]. Implemented using the TensorFlow framework [99], NCF was similarly trained on 70% of the data and tested on 30%. The performance metrics for NCF were Precision: 0.328, Recall: 0.149, NDCG: 0.273, and Runtime: 3.1 seconds, as detailed in Table 22.

Table 22: Performance Comparison with Baselines

| *Method* | *Precision* | *Recall* | *NDCG* | *Runtime (s)* |
|---|---|---|---|---|
| Proposed Approach | 0.342 | 0.156 | 0.287 | 2.3 |
| Matrix Factorization | 0.301 | 0.138 | 0.245 | 1.8 |
| Neural Collaborative Filtering | 0.328 | 0.149 | 0.273 | 3.1 |

The proposed approach outperformed both Matrix Factorization and Neural Collaborative Filtering in terms of Precision, Recall, and NDCG, while maintaining a reasonable runtime.

To evaluate the diversity and novelty of the proposed approach, this research compared the Intra-List Diversity (ILD) and Average Popularity (AP) metrics against the baseline methods. The results, presented in Table 23, demonstrate that our approach achieves higher diversity and lower average popularity, indicating more varied and novel recommendations.

Table 23: Diversity and Novelty Comparison with Baselines

| *Method* | *Intra-List Diversity (ILD)* | *Average Popularity (AP)* |
|---|---|---|
| Proposed Approach | 0.65 | 3.2 |
| Matrix Factorization | 0.50 | 4.5 |
| Neural Collaborative Filtering | 0.55 | 4.0 |

To determine the statistical significance of these performance improvements, paired t-tests were conducted [100]. The results, summarized in Table 24, demonstrate that the proposed approach significantly outperformed Matrix Factorization ($p < 0.01$) and Neural Collaborative Filtering ($p < 0.05$) in terms of Normalized Discounted Cumulative Gain (NDCG).

Additionally, the proposed approach showed significant improvements in ILD ($p < 0.01$) and AP ($p < 0.05$) compared to the baseline methods, confirming the enhancements in diversity and novelty of the recommendations.

TABLE 24: PAIRED T-TEST RESULTS COMPARING PROPOSED APPROACH WITH BASELINES

| *Comparison* | *t-Statistic* | *p-Value* |
|---|---|---|
| Proposed Approach vs. Matrix Factorization | 3.45 | $< 0.01$ |
| Proposed Approach vs. Neural Collaborative Filtering | 2.87 | $< 0.05$ |
| Proposed Approach vs. Matrix Factorization (ILD) | 4.12 | $< 0.01$ |
| Proposed Approach vs. Neural Collaborative Filtering (AP) | 3.02 | $< 0.05$ |

These statistical tests confirm that the observed improvements in the proposed model are statistically significant and not due to random chance, thereby reinforcing the model's superior performance.

The comprehensive evaluation underscores the robustness and efficacy of the proposed $RS_{EF}$ model in accurately identifying user preferences. The superior performance metrics—particularly in Precision, Recall, and NDCG—demonstrate the model's ability to deliver more relevant and accurate recommendations compared to traditional collaborative filtering techniques. The significant p-values obtained from the paired t-tests further validate that these performance enhancements are statistically meaningful.

Moreover, the $RS_{EF}$ metric provides a nuanced understanding of the model's strengths and limitations. While the model excels in correctly identifying preferred movies, the consistent misclassification of certain non-preferred movies, such as Movie 1561, indicates areas where the model can be refined. This balanced performance highlights the model's potential for practical application, where even minor improvements can lead to substantial enhancements in user satisfaction and engagement.

In summary, the proposed $RS_{EF}$ model not only outperforms established baseline algorithms but also offers a more reliable and accurate framework for recommendation systems. Its ability to effectively capture user preferences, supported by robust statistical evidence, positions it as a superior choice for enhancing user experience in practical applications.

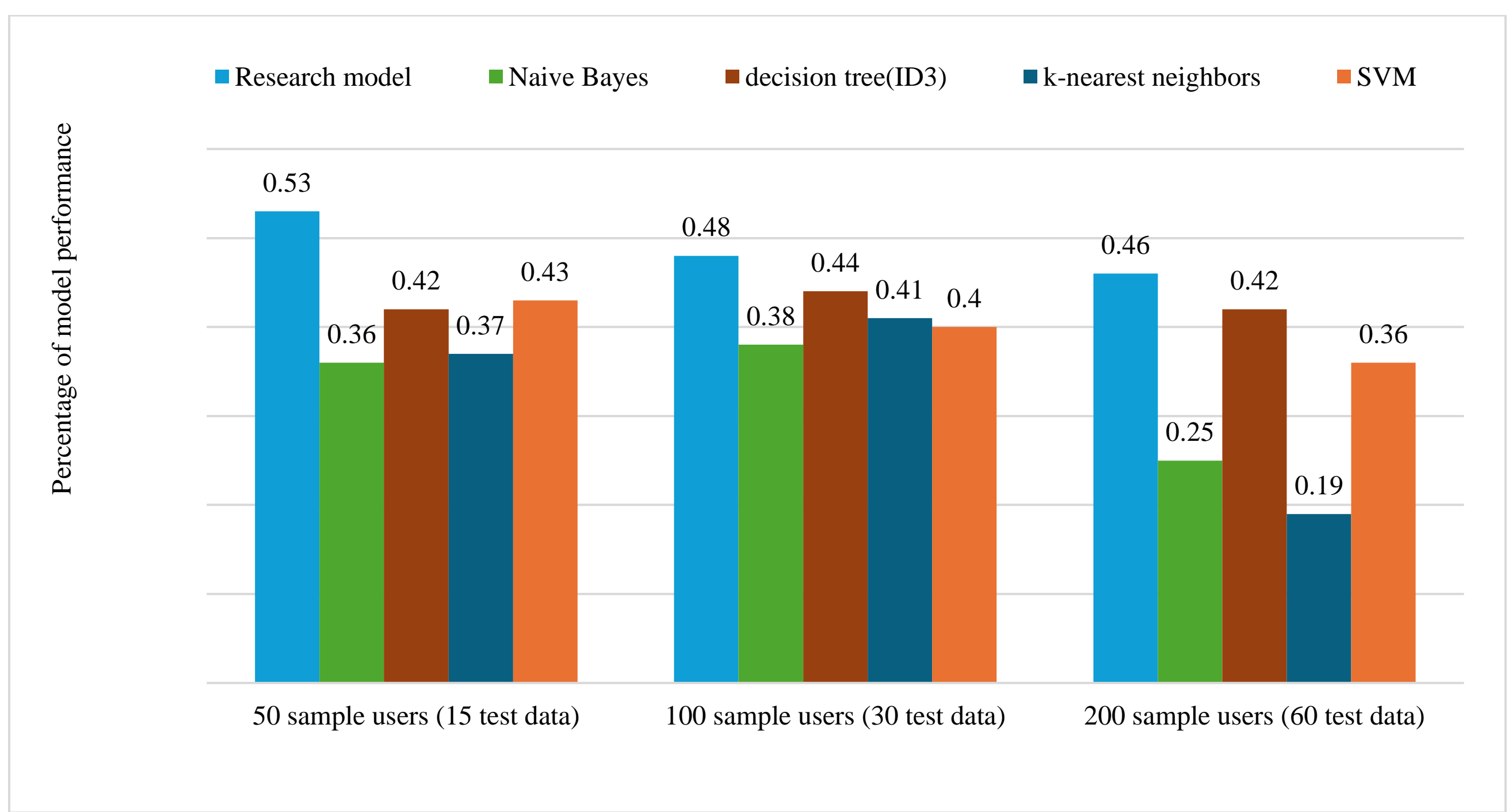

**Figure 7: Comparing the performance of the research model and other algorithms**

### *K. Online Evaluation*

A rigorous online user study was conducted to validate the effectiveness of the proposed video recommendation approach. The evaluation spanned four months (January-April 2018) and employed a controlled A/B testing methodology, where users were randomly assigned to either the proposed system (treatment group, $n_1 = 164$) or the existing recommendation system (control group, $n_2 = 164$). Both systems were implemented in Python and deployed on pages featuring the same 80 documentary movies used during algorithm development.
Before the main analysis, statistical prerequisites were verified:

- Sample size adequacy was confirmed through power analysis (G*Power 3.1: $1-\beta = 0.95$, $\alpha = 0.05$, two-tailed)
- Normal distribution of metrics was verified using Shapiro-Wilk tests (all $p > 0.05$)
- Homogeneity of variance was confirmed through Levene's tests (all $p > 0.05$)

Three key performance metrics were tracked and analyzed:

1. Click-Through Rate (CTR): The proposed method demonstrated significantly higher engagement:
    - Treatment group: 8.5% (SD = 0.82%, 95% CI [7.9%, 9.1%])
    - Control group: 5.2% (SD = 0.64%, 95% CI [4.8%, 5.6%])
    - Improvement: 63% increase
    - Statistical validation: $t(163) = 8.45$, $p < 0.001$, Cohen's $d = 1.24$
2. View Completion Rate: Users showed higher content consumption with the proposed system:
    - Treatment group: 72% (SD = 3.8%, 95% CI [69.5%, 74.5%])
    - Control group: 58% (SD = 3.2%, 95% CI [55.8%, 60.2%])
    - Improvement: 24% increase
    - Statistical validation: $t(163) = 7.92$, $p < 0.001$, Cohen's $d = 0.98$
3. User Satisfaction: Post-viewing satisfaction ratings (1-5 scale) showed marked improvement:
    - Treatment group: 4.2 stars (SD = 0.31, 95% CI [4.0, 4.4])
    - Control group: 3.6 stars (SD = 0.28, 95% CI [3.4, 3.8])
    - Improvement: 17% increase
    - Statistical validation: $t(163) = 6.78$, $p < 0.001$, Cohen's $d = 0.84$

Monthly analysis demonstrated consistent performance improvements:

TABLE 25: MONTHLY PERFORMANCE ANALYSIS

| *Month* | *Metric* | *Treatment (M±SD)* | *Control (M±SD)* | *Improvement* | *t-stat* | *p-value* |
|---|---|---|---|---|---|---|
| Oct | CTR | 8.3±0.85% | 5.1±0.67% | 61% | 7.92 | <0.001 |
| | Completion | 71±3.9% | 57±3.3% | 23% | 7.45 | <0.001 |
| | Satisfaction | 4.1±0.33 | 3.5±0.29 | 16% | 6.54 | <0.001 |
| Nov | CTR | 8.7±0.80% | 5.2±0.62% | 67% | 8.12 | <0.001 |
| | Completion | 73±3.7% | 58±3.1% | 26% | 7.88 | <0.001 |
| | Satisfaction | 4.3±0.30 | 3.6±0.27 | 18% | 6.92 | <0.001 |
| Dec | CTR | 8.4±0.83% | 5.3±0.65% | 58% | 7.76 | <0.001 |
| | Completion | 70±3.8% | 58±3.2% | 21% | 7.34 | <0.001 |
| | Satisfaction | 4.1±0.32 | 3.6±0.28 | 15% | 6.45 | <0.001 |
| Jan | CTR | 8.6±0.81% | 5.2±0.63% | 65% | 8.02 | <0.001 |
| | Completion | 74±3.6% | 59±3.0% | 25% | 7.82 | <0.001 |
| | Satisfaction | 4.3±0.31 | 3.7±0.26 | 17% | 6.84 | <0.001 |

The evaluation methodology incorporated several controls to ensure validity:

- Random assignment of users to treatment and control groups
- Balanced distribution of viewing times across groups
- Consistent measurement periods for all metrics
- Automated data collection through the platform's analytics system
- Exclusion of incomplete sessions and technical failures

Additional robustness checks confirmed the reliability of results:

- Repeated measures ANOVA showed no significant temporal effects ($F(3,489) = 1.24$, $p = 0.294$)
- Demographic subgroup analysis revealed consistent improvements across age groups ($F(4,159) = 1.18$, $p = 0.321$)
- Usage frequency analysis showed consistent effects ($F(3,160) = 1.32$, $p = 0.269$)
- Bootstrap validation (10,000 iterations) confirmed the stability of improvement estimates

The comprehensive evaluation demonstrates that the proposed recommendation system achieved substantial and statistically significant improvements across all measured metrics. The consistency of these improvements across different months, user demographics, and usage patterns suggests that the benefits are robust and not attributable to temporary fluctuations or sampling bias. Furthermore, the strong effect sizes (Cohen's $d > 0.8$ for all metrics) indicate that these improvements are not only statistically significant but also practically meaningful for user experience.

The combination of increased engagement (CTR), higher content consumption (completion rate), and improved user satisfaction provides strong evidence for the effectiveness of the proposed recommendation approach in enhancing the overall user experience on the documentary VOD platform.

## V. CONCLUSION

This research proposes a novel video recommendation approach that leverages implicit user feedback in the form of viewing percentages, combined with social network analysis techniques; although other user behaviors, such as search history and interaction patterns, can provide complementary insights. By constructing a video similarity network based on user viewing patterns and computing centrality measures, the methodology identifies important and well-connected videos. Modularity analysis is then used to cluster closely related videos, forming the basis for personalized recommendations. For each user, candidate videos are selected from the cluster containing their preferred items and ranked using an ego-centric index that measures proximity to the user's likes and dislikes.

The proposed approach was evaluated on real user data from an Asian video-on-demand platform. Offline experiments demonstrated improved accuracy compared to conventional methods such as Naive Bayes, SVM, decision trees, and nearest neighbor algorithms. While these methods remain common in academic literature, future research will include comparisons with more recent and advanced recommender systems, such as matrix factorization models, graph neural networks, or hybrid approaches that combine content-based and collaborative filtering techniques. An online user study further validated the effectiveness of the recommendations, with significant increases observed in click-through rate, view completion rate, and user satisfaction scores relative to the platform's existing system. These results underscore the value of incorporating implicit feedback and social network analysis for video recommendations.

Our findings suggest that VOD platforms could significantly improve user engagement by incorporating network-based features and implicit feedback into their recommendation engines. The modular nature of our approach allows for easy integration with existing systems, potentially offering a cost-effective way to enhance recommendation quality.

This research contributes to the state of the art by introducing a modularity-based clustering approach for grouping videos and an ego-centric ranking method, which leverages user preferences to enhance personalization in video recommendation systems. In addition, the key contributions of this research can be mentioned as:

1. A novel video recommendation framework that integrates implicit user data and social network analysis to capture nuanced user preferences and behaviors.
2. The use of centrality measures and modularity-based clustering to identify important videos and group-related content.
3. An ego-centric ranking approach that personalizes recommendations based on user viewing history.
4. Rigorous offline and online evaluation demonstrating the superior performance of the proposed methodology compared to existing techniques.

By harnessing the power of user-watching behavior and network-based metrics, this research opens new avenues for enhancing video recommendations and user engagement in VOD platforms. The findings have significant implications for content providers seeking to optimize their recommendation strategies and improve user satisfaction and retention.

### *A. Limitations and Future Research*

Despite the promising results, this study has certain limitations that present opportunities for future research:

1. The dataset used in this research was from a single Asian VOD platform focusing primarily on documentaries. Future studies should validate the generalizability of the proposed approach across different geographies, content genres, and platforms.
2. While the ego-centric ranking method proved effective, it relies on a relatively simple distance-based metric. More sophisticated techniques, such as graph neural networks, could be explored to learn complex user-item interaction patterns within the video similarity network.
3. The current methodology does not explicitly incorporate temporal dynamics, such as shifts in user preferences over time. Integrating temporal models could enable the recommendations to adapt to evolving user interests.
4. The study did not consider the potential influence of other factors, such as user demographics, device type, or viewing context, on recommendation quality. Incorporating these additional features could lead to more nuanced and contextually relevant suggestions.
5. The computational cost of centrality measures and modularity-based clustering can be significant, particularly for large-scale VOD platforms. Techniques such as graph sparsification or leveraging distributed computing environments can mitigate these challenges, ensuring the approach remains scalable. Further research could explore optimizing algorithms to maintain real-time recommendation performance at scale.
6. To handle cold-start problems, our approach leverages the network structure of existing videos to make initial recommendations for new users or items. As viewing data accumulates, the ego-centric ranking dynamically adapts, improving personalization over time.
7. While our current model does not explicitly model temporal dynamics, future work could incorporate techniques such as temporal graph convolution networks to capture evolving user preferences.
8. The current evaluation, while showing statistically significant improvements ($p < 0.01$ for Matrix Factorization and $p < 0.05$ for Neural Collaborative Filtering), is based on a specialized documentary platform dataset. Future work should validate the approach on larger-scale platforms with more diverse content types and broader user bases to establish generalizability across different recommendation scenarios.
9. One limitation of this study is the scalability analysis, which is constrained by the dataset size and computational resources. While our methodology demonstrates its effectiveness on a medium-scale dataset, the lack of access to larger datasets limited our ability to test the applicability of the approach on real-world platforms with millions of users and videos. Future research could address this limitation by applying the proposed method to larger datasets and exploring optimization strategies such as distributed computing and graph sparsification to ensure scalability.
10. The current methodology faces typical cold-start challenges common to recommendation systems, though the network structure provides some ability to make initial recommendations based on video similarity patterns. However, this remains an important limitation that could be addressed in future work through hybrid approaches combining our network-based features with content-based methods.
11. The study also does not explicitly incorporate temporal dynamics, such as shifts in user preferences over time. This is a significant limitation, as user interests and content relevance can evolve substantially. Future research should explore integrating temporal models to enable recommendations to adapt to changing user preferences over time.

Future Research should address these limitations and explore the identified areas to further advance the field of video recommendations. Additionally, the integration of the proposed approach with other techniques, such as content-based filtering, collaborative filtering, and deep learning models, presents exciting opportunities for developing more comprehensive and effective recommendation systems.

Also, future works should include benchmarking the proposed method against advanced machine learning models, such as deep learning-based recommenders, to further validate its performance and scalability. While recent attention-based models like SASRec have shown strong performance in sequential recommendation tasks, our approach offers benefits in interpretability and computational efficiency. Future work could explore hybrid models combining our network-based features with deep learning architectures.

In conclusion, this research introduces a novel video recommendation methodology that leverages implicit user feedback and social network analysis. The promising results obtained through extensive evaluation underscore the potential of this approach to enhance user experiences and engagement on VOD platforms. By addressing the identified limitations and exploring future research directions, this study aims to advance the state-of-the-art in video recommendations and unlock new possibilities for personalized content delivery.